\documentclass[aps,prl,reprint,amsmath,amssymb]{revtex4-2}
\usepackage{graphicx}
\usepackage{bm}
\begin{document}
\raggedbottom
\title{Probing spin dynamics in MnSi under pressure by energy-integrated time-of-flight small-angle neutron scattering}
\author{E. V. Altynbaev}\affiliation{Saint Petersburg State University, Saint Petersburg, Russia}\affiliation{Institute for High Pressure Physics, Russian Academy of Sciences, Troitsk, Moscow, Russia}
\author{D. O. Skanchenko}\affiliation{Mozhaisky Military Space Academy, Saint Petersburg, Russia}\affiliation{Institute for High Pressure Physics, Russian Academy of Sciences, Troitsk, Moscow, Russia}
\author{Zhenhua Xie}\affiliation{Institute of High Energy Physics, Chinese Academy of Sciences, Beijing 100049, China; Dongguan Neutron Science Center, Dongguan 523803, China}
\author{Yubin Ke}\affiliation{Institute of High Energy Physics, Chinese Academy of Sciences, Beijing 100049, China; Dongguan Neutron Science Center, Dongguan 523803, China}
\author{Yuan Bao}\affiliation{Institute of High Energy Physics, Chinese Academy of Sciences, Beijing 100049, China; Dongguan Neutron Science Center, Dongguan 523803, China}
\author{Hui Cheng}\affiliation{Institute of High Energy Physics, Chinese Academy of Sciences, Beijing 100049, China; Dongguan Neutron Science Center, Dongguan 523803, China}
\author{A. V. Tsvyashchenko}\affiliation{Institute for High Pressure Physics, Russian Academy of Sciences, Troitsk, Moscow, Russia}
\date{\today}
\begin{abstract}
We demonstrate that energy-integrated time-of-flight small-angle neutron scattering can determine the long-wavelength spin-wave stiffness of field-polarized MnSi inside a pressure cell. A forward model maps neutron energy loss and gain into the recorded time-of-flight coordinates, allowing a joint fit of radial profiles in several apparent-wavelength windows to constrain a common stiffness. Ambient-pressure powder and polycrystal measurements reproduce the established stiffness scale of MnSi. At effective pressures of 3.5 and 7.8 kbar, the fitted stiffness spans 43--61 meV\,\AA$^2$ across the measured states and is comparable in magnitude to the scale inferred from the helical wave vector and applied upper critical field. The characteristic SANS ordering temperature decreases from 28.8 K at ambient pressure to 17.4 K at 7.8 kbar. The long-wavelength gradient stiffness alone therefore does not account for the pressure evolution of helimagnetic ordering. These results demonstrate a route to quantitative studies of spin-wave dynamics under pressure on time-of-flight SANS instruments without final-energy analysis.
\end{abstract}
\maketitle

MnSi is an itinerant chiral magnet whose competing exchange and Dzyaloshinskii--Moriya interactions produce long-period helical order below approximately 29 K \cite{ish76,bak,grig06,muehl,adams}. Hydrostatic pressure strongly suppresses this ordering scale and changes the magnetic state \cite{pf97,pf04,fak,pf07,bannen}. Static ordering temperatures and phase boundaries characterize the stability of magnetic order, whereas dynamical measurements reveal the energies of its excitations. Determining how these energies evolve under pressure is therefore essential to understanding the suppression of helimagnetism.

The spin-wave stiffness $A$ sets the low-$q$ curvature of the dispersion and characterizes the energy cost of long-wavelength magnetic variations. Inelastic-neutron measurements have resolved spin waves, helimagnon bands, and nonreciprocal excitations in MnSi \cite{ish77,jan,kug,weber18,weber19}. Recent zero-field measurements relate the pressure evolution of the ordered moment $m(T)$ to an exchange-related scale \cite{DeReotier2024}, motivating a direct determination of the dynamical stiffness. The central question is whether the strong reduction of the ordering temperature is accompanied by a comparable reduction of the long-wavelength excitation scale.

Earlier small-angle neutron-scattering studies demonstrated stiffness extraction in field-polarized MnSi and polycrystalline FeGe \cite{grig15,siegfried17}. Wavelength-resolved time-of-flight SANS (TOF-SANS) offers an opportunity to extend this approach to the restrictive pressure-cell environment. The weak inelastic signal must be distinguished from the cell background, and its interpretation must account for the recorded TOF coordinates: neutron energy transfer changes the arrival time and hence the wavelength and momentum transfer assigned by elastic reduction \cite{seeger,du,campillo}.

Here we use ambient powder and polycrystal measurements as benchmarks and determine the dynamical stiffness at two effective pressures. A forward treatment of the wavelength-dependent profiles is compared with a stiffness scale constructed from structural SANS measurements. The resulting long-wavelength stiffness remains substantial and on a scale comparable to the structural estimate, while the characteristic temperature of stable helimagnetic order falls much more strongly.

\begin{figure*}[t]\centering\includegraphics[width=\textwidth]{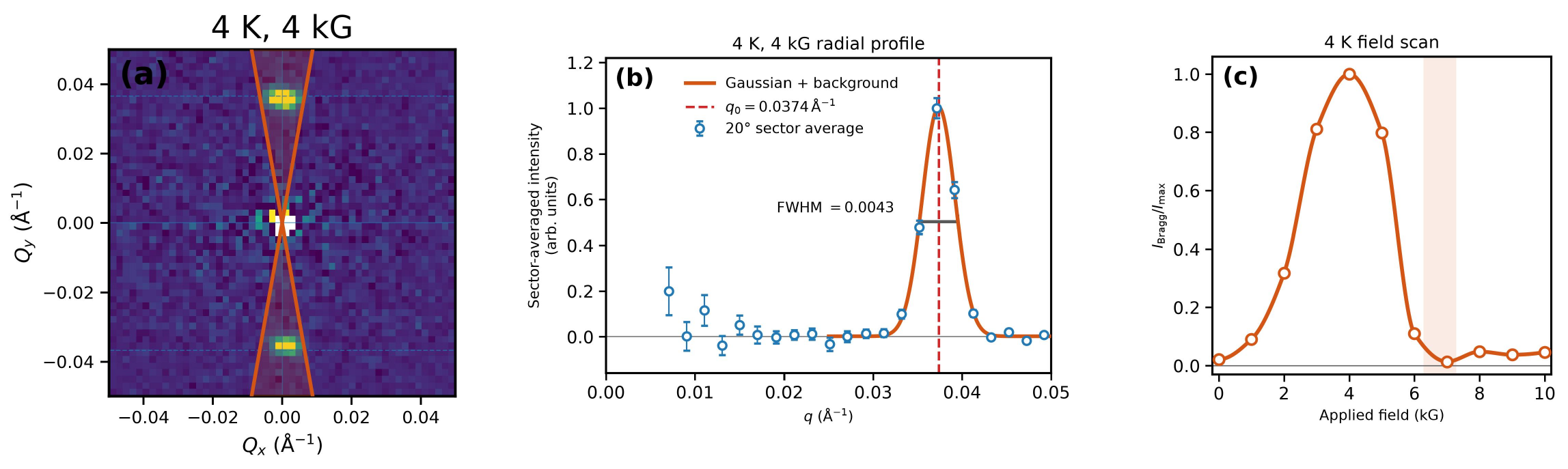}
\caption{Structural SANS at $p_{\rm eff}=7.8\pm0.9$ kbar and 4 K. (a) Magnetic pattern at 4 kG with the two $20^\circ$ sectors used for radial averaging. (b) Corresponding radial profile, with peak center $q_0=0.0374$ \AA$^{-1}$ and observed FWHM $0.0043$ \AA$^{-1}$. (c) Integrated Bragg intensity versus applied field; the band marks $H_{C2}^{\rm appl}=6.78\pm0.50$ kG. Lines describe the peak in (b) and guide the eye in (c).}\label{fig:struct}\end{figure*}
\begin{figure*}[t]\centering\includegraphics[width=\textwidth]{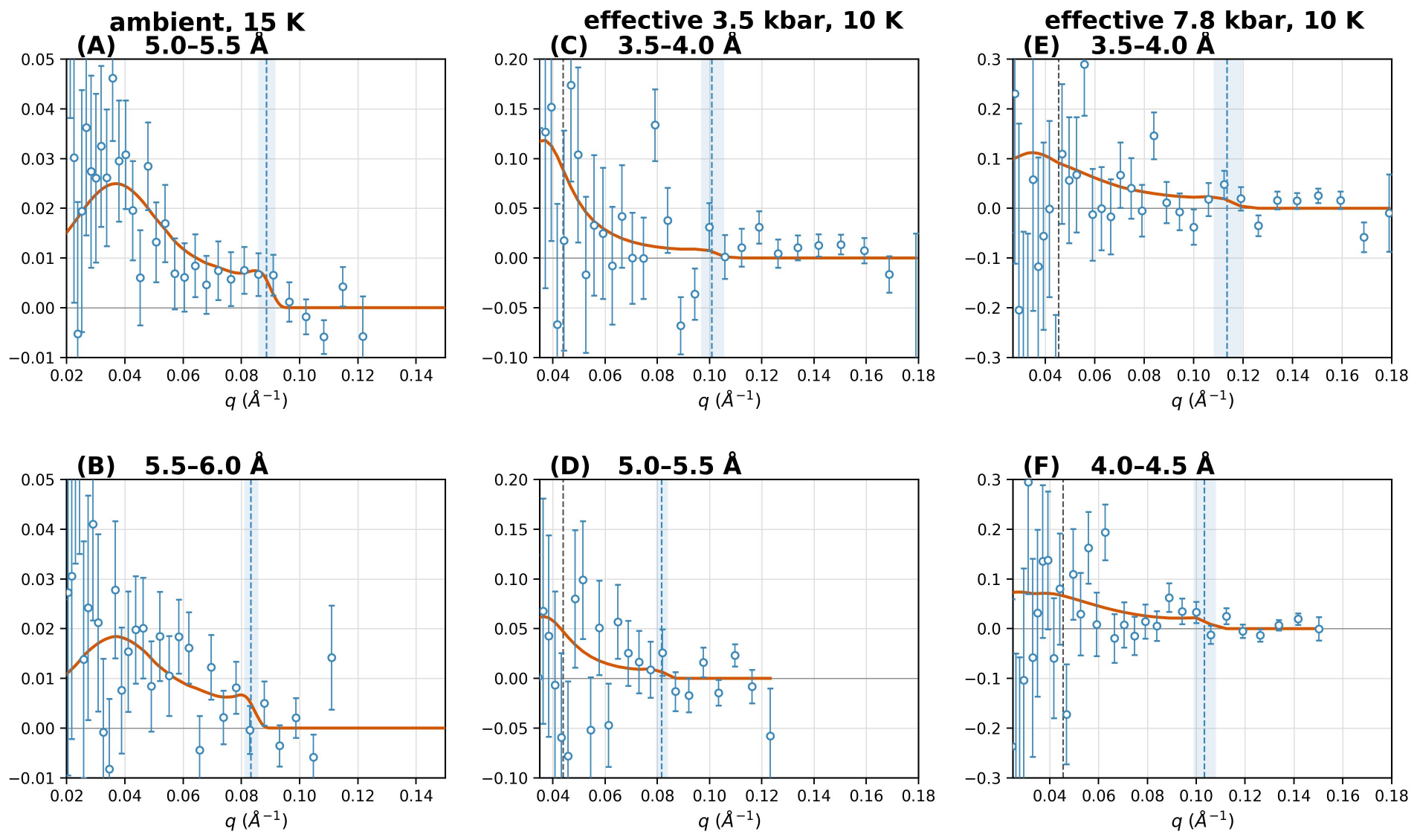}
\caption{Wavelength-resolved TOF-SANS profiles (symbols with statistical errors) and calculated profiles (solid lines). Columns show ambient powder at 15 K and 8 kG, $p_{\rm eff}=3.5\pm0.8$ kbar at 10 K and 7 kG, and $p_{\rm eff}=7.8\pm0.9$ kbar at 10 K and 9 kG. Each panel is labelled with its recorded apparent-wavelength interval. Each state has one common $A_{\rm dyn}$ and separate intensity factors for its wavelength windows. Dashed blue lines mark the calculated $q_{\rm app}$ position of a characteristic profile feature after TOF mapping at the central wavelength of each interval; pale-blue bands span its calculated positions across that wavelength interval.}\label{fig:profiles}\end{figure*}
\begin{figure*}[t]\centering\includegraphics[width=\textwidth]{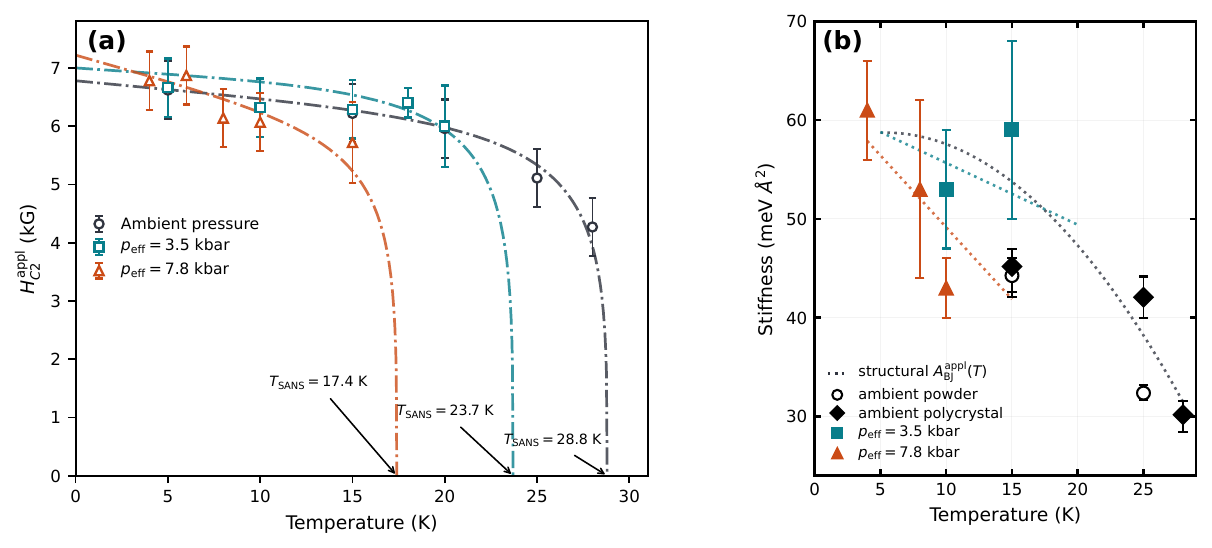}
  \caption{Temperature-dependent magnetic scales. (a) Measured applied upper critical fields $H_{C2}^{\rm appl}$ and smooth guides ending at the corresponding $T_{\rm SANS}$ values, marked by arrows. (b) Unconnected dynamical-stiffness points $A_{\rm dyn}$ and smooth guides to the structural stiffness $A_{\rm BJ}^{\rm appl}(T)$. Open black circles denote ambient powder, filled black diamonds denote ambient polycrystal, and teal squares and orange triangles denote the $p_{\rm eff}=3.5$ and 7.8 kbar series.}\label{fig:compare}\end{figure*}
\section{Structural reference under pressure}

Structural SANS establishes the magnetic state and provides a stiffness comparison scale through the helical wave vector $k_s(T)$ and applied upper critical field $H_{C2}^{\rm appl}(T)$. Measurements were performed on BL01 at the China Spallation Neutron Source, with the pressure sample in a CuBe piston-cylinder cell \cite{ke,sm}. Temperature scans of helical Bragg intensity locate the characteristic SANS ordering temperature $T_{\rm SANS}$, defined by the disappearance of that scattering on warming. Its values are $28.8\pm0.5$ K at ambient pressure and $23.7\pm1.1$ and $17.4\pm1.3$ K for the two pressure loadings. Calibration against the established low-pressure ordering dependence assigns effective pressures $p_{\rm eff}=3.5\pm0.8$ and $7.8\pm0.9$ kbar \cite{sm}.

Bragg-peak positions determine the temperature-dependent $k_s$, while field scans of integrated Bragg intensity determine $H_{C2}^{\rm appl}$. Figure~\ref{fig:struct} shows representative magnetic scattering, its radial profile, and the field dependence at the higher pressure. In the continuum helimagnet description, $k_s$ sets the modulation length and $H_{C2}$ sets the field scale required to unwind the helix into the polarized state. Their combination gives the spin-wave stiffness associated with the quadratic long-wavelength gradient term \cite{bak,kataoka,maleyev}:
\begin{equation}\label{eq:bj}
A_{\rm BJ}^{\rm appl}=\frac{g\mu_BH_{C2}^{\rm appl}}{k_s^2},\qquad g=2.
\end{equation}
We use the applied-field convention consistently. The dynamical analysis uses the measured $k_s$ and $H_{C2}^{\rm appl}$ as state-specific inputs, while $A_{\rm dyn}$ is determined from the wavelength-dependent profiles.

\section{The recorded spin-wave profile}

In conventional energy-integrated TOF-SANS, the detector records neutron arrival time and position while integrating over final neutron energies. Inelastic scattering exchanges energy between neutrons and spin-wave excitations, so a scattered neutron leaves with a velocity different from that assumed in elastic reduction. Its arrival time is therefore assigned an apparent wavelength and momentum transfer that depend on the energy transfer. This gives the spin-wave contribution a wavelength-dependent signature in the radial profile. Several recorded wavelength intervals probe the same dispersion through different TOF mappings, allowing their combined shapes to constrain a common $A_{\rm dyn}$. We refer to this contribution as spin-wave small-angle neutron scattering (SWSANS).

The polarized-state dispersion is represented by quadratic modes shifted by the helical wave vector,
\begin{equation}\label{eq:disp}
\begin{aligned}
\epsilon_s(\bm Q)&=A_{\rm dyn}|\bm Q-s\bm k_s|^2+\Delta,\\
\Delta&=g\mu_B\Delta H,\\
\Delta H&=H_{\rm appl}-H_{C2}^{\rm appl},\qquad s=\pm1.
\end{aligned}
\end{equation}
Here $\bm Q=\bm k_i-\bm k_f$. The index $s$ labels the two wave-vector shifts, separately from neutron energy loss or gain. Energy conservation, $E_i-E_f=\pm\epsilon_s$, determines the final wave vector. Loss and gain carry the Bose weights $n(\epsilon_s)+1$ and $n(\epsilon_s)$, respectively, with $n(\epsilon)=[\exp(\epsilon/k_BT)-1]^{-1}$; the cross section also includes the magnetic polarization factor and $k_f/k_i$.

For incident path $L_1$ and the angle-dependent detector path $L_{2,\rm eff}(\theta)$ defined from the nominal BL01 geometry in the Supplemental Material, elastic reduction assigns
\begin{equation}\label{eq:tof}
\begin{aligned}
\lambda_{\rm app}&=\frac{L_1\lambda_i+L_{2,\rm eff}\lambda_f}{L_1+L_{2,\rm eff}},\\
q_{\rm app}&=\frac{4\pi}{\lambda_{\rm app}}\sin\frac{\theta}{2},
\end{aligned}
\end{equation}
where $\theta$ is the scattering angle and $\lambda_{i,f}=2\pi/k_{i,f}$. The calculation projects the inelastic cross section into these recorded coordinates using the normalized BL01 incident-wavelength/source weighting, state-specific $k_s$ and $H_{C2}^{\rm appl}$, detector and sector averaging, momentum resolution, and the recorded wavelength windows. The complete calculation and background treatment are given in Supplemental Material \cite{sm}.

Independent intensity factors $C_i$ allow the scale of each wavelength window to vary, whereas $A_{\rm dyn}$ controls the kinematics and wavelength-dependent profile morphology. Figure~\ref{fig:profiles} illustrates the resulting determination: the measured profiles evolve systematically with wavelength, and the corresponding calculated features shift through the TOF mapping. One common stiffness describes the informative windows of each state. The pressure-cell profiles retain this wavelength-dependent information, allowing the same measurement principle to determine $A_{\rm dyn}$ under pressure.

\section{Dynamic stiffness and structural comparison}

Ambient powder and polycrystal measurements benchmark the approach across preparation and temperature [Fig.~\ref{fig:compare}(b)]. At 15 K, they give $44\pm2$ and $45\pm3$ meV\,\AA$^2$, respectively. These ambient values reproduce the established MnSi stiffness scale and its overall temperature dependence \cite{grig15}. We apply the same profile-based procedure under pressure, using the corresponding state-specific structural $k_s$ and $H_{C2}^{\rm appl}$. The higher-temperature measurements resolve a reduction of stiffness while remaining on a comparable tens-of-meV\,\AA$^2$ scale. Both preparations exhibit this characteristic scale and its temperature evolution. The measured states and corresponding structural comparisons are listed in Supplemental Table~S1 \cite{sm}.

Under pressure, the stiffness remains near 50--60 meV\,\AA$^2$ at $p_{\rm eff}=3.5\pm0.8$ kbar: $A_{\rm dyn}=53\pm6$ and $59\pm9$ meV\,\AA$^2$ at 10 and 15 K. At $p_{\rm eff}=7.8\pm0.9$ kbar it evolves from about 60 to about 40 meV\,\AA$^2$ over the measured temperature range, with values $61\pm5$, $53\pm9$, and $43\pm3$ meV\,\AA$^2$ at 4, 8, and 10 K. These points remain on a scale comparable to the structural stiffness curves in Fig.~\ref{fig:compare}(b). In contrast, the ordering temperatures marked in Fig.~\ref{fig:compare}(a) shift strongly downward with pressure. The measured low-$q$ stiffness thus remains substantial as stable helimagnetic order is suppressed.

\section{Discussion}

The two observables characterize different aspects of magnetism. The stiffness $A$ measures the low-$q$ curvature of the spin-wave dispersion and the energy cost of long-wavelength spatial variations in the measured magnetic state. The ordering locator $T_{\rm SANS}$ characterizes the temperature stability of long-range helimagnetic order. Both reflect magnetic interactions, but a single low-$q$ coefficient does not specify the full magnetic free energy governing thermal ordering. Their relation therefore need not be a universal one-to-one proportionality. Effective-exchange calculations for metallic ferromagnets illustrate this distinction: the spin-wave stiffness is set by the small-$q$ curvature, whereas mean-field and random-phase estimates of $T_C$ involve averages over the spin-wave spectrum \cite{pajda01}. In the investigated pressure range, the strong reduction of $T_{\rm SANS}$ without a corresponding stiffness collapse shows that the pressure evolution of order contains physics beyond this gradient coefficient.

The structural comparison places this observation on a consistent magnetic energy scale. The combination of $k_s(T)$ and $H_{C2}^{\rm appl}(T)$ describes the field and modulation scales of the helix, whereas the TOF-SANS profiles determine its dynamical stiffness through inelastic kinematics. Their comparable magnitudes and overall temperature evolution support the physical consistency of the extracted stiffness. The offsets between individual ambient estimates show the appropriate scope of this comparison: $A_{\rm BJ}^{\rm appl}$ supplies a characteristic scale, while $A_{\rm dyn}$ is determined from the wavelength-dependent profiles.

Dalmas de R\'eotier et al. inferred a pressure-dependent exchange parameter $J$ from the zero-field thermal evolution of the ordered moment near the critical pressure \cite{DeReotier2024}. Our finite-field, low-$q$ stiffness probes a different response at moderate pressures. Relating these quantities requires the respective magnetic states and model definitions to be taken into account; the studies together motivate a description of the pressure evolution that distinguishes ordering and excitation scales.

Recent anisotropy and damping results likewise emphasize the scope of an effective scalar stiffness \cite{khoroshiy26,utesov26}. Here $A_{\rm dyn}$ describes an effective scalar, orientationally averaged response in the field-defined geometry; the present measurements do not identify the microscopic origin of the remaining pressure evolution.

Wavelength-resolved, energy-integrated TOF-SANS thus provides access to the dynamical stiffness of MnSi inside a pressure cell. The measured stiffness remains substantial and comparable in magnitude to the structurally inferred scale, while $T_{\rm SANS}$ is strongly reduced. The pressure evolution of helimagnetic ordering is therefore not captured by the long-wavelength gradient stiffness alone.
\begin{acknowledgments}
\noindent\begin{minipage}{\columnwidth}
\noindent\textbf{ACKNOWLEDGMENTS}\par
We thank the China Spallation Neutron Source (CSNS) for beamtime on BL01 and the BL01 instrument team and CSNS staff for help preparing and conducting the experiment.
This research was funded by the Russian Science Foundation,
Grant No. 25-12-68013 \mbox{(22-12-00008-$\pi$)}.
\end{minipage}\par
\end{acknowledgments}

\end{document}